\documentclass[fleqn,usenatbib,useAMS]{mnras}

\DeclareRobustCommand{\VAN}[3]{#2}
\let\VANthebibliography\thebibliography
\def\thebibliography{\DeclareRobustCommand{\VAN}[3]{##3}\VANthebibliography}

\usepackage{amsmath,amsfonts,amssymb} 
\usepackage{bm} 
\usepackage[table,  dvipsnames]{xcolor} 
\usepackage{hyperref} 
\usepackage{nicefrac} 
\usepackage{placeins}

\usepackage[varg]{txfonts} 
\usepackage[T1]{fontenc}

\hypersetup{
	pdftitle = {Buoyancy response of a disk to an embedded planet: a cross-code comparison at high resolution},
	pdfauthor = {Alexandros Ziampras, Sijme-Jan Paardekooper, Richard P. Nelson},
	pdfsubject = {},
	colorlinks = {true},
	linkcolor = [rgb]{0,0.35,0.7},
	citecolor = [rgb]{0,0.35,0.7},
	filecolor = [rgb]{0.61,0,0},
	urlcolor = [rgb]{0.61,0,0},
}

\usepackage{graphicx}  
\graphicspath{{figures/}}

\newcommand{\DP}[2]{\frac{\partial{#1}}{\partial{#2}}}
\newcommand{\D}[2]{\frac{\text{d}{#1}}{\text{d}{#2}}}

\newcommand{\G}{\text{G}}
\newcommand{\Mstar}{M_\star}

\newcommand{\Rp}{R_\mathrm{p}}
\newcommand{\Mp}{M_\mathrm{p}}
\newcommand{\hp}{h_\mathrm{p}}

\newcommand{\Pp}{P_\mathrm{p}}

\newcommand{\Msun}{\mathrm{M}_\odot}

\newcommand{\Rgas}{\mathcal{R}}
\newcommand{\cs}{c_\mathrm{s}}
\newcommand{\csiso}{c_\mathrm{s,iso}}

\newcommand{\OmegaK}{\Omega_\mathrm{K}}

\newcommand{\rhomid}{\rho_\mathrm{mid}}

\newcommand{\vel}{\bm{u}}

\newcommand{\xh}{{x}_\mathrm{h}}

\defcitealias{mcnally-etal-2020}{MN20}
\newcommand{\mcnally}{\citetalias{mcnally-etal-2020}}
\newcommand{\mcnallyp}{\citepalias{mcnally-etal-2020}}

\title[Planet-driven buoyancy waves: a code comparison]{Buoyancy response of a disk to an embedded planet:\\a cross-code comparison at high resolution}

\author[A.~Ziampras et al.]{
Alexandros~Ziampras$^{1}$\thanks{E-mail: a.ziampras@qmul.ac.uk},
Sijme-Jan~Paardekooper$^{1,2}$,
Richard~P.~Nelson$^{1}$
\\
$^{1}$Astronomy Unit, Department of Physics and Astronomy, Queen Mary University of London, London E1 4NS, UK\\
$^{2}$Faculty of Aerospace Engineering, Delft University of Technology, Delft, The Netherlands\\
}

\date{Accepted XXX. Received YYY; in original form ZZZ}
\pubyear{2023}

\begin{document}
\label{firstpage}
\pagerange{\pageref{firstpage}--\pageref{lastpage}}
\maketitle

\begin{abstract}
	In radiatively inefficient, laminar protoplanetary disks, embedded planets can excite a buoyancy response as gas gets deflected vertically near the planet. This results in vertical oscillations that drive a vortensity growth in the planet's corotating region, speeding up inward migration in the type-I regime. We present a comparison between \texttt{PLUTO}/\texttt{IDEFIX} and \texttt{FARGO3D} using 3D, inviscid, adiabatic numerical simulations of planet--disk interaction that feature the buoyancy response of the disk, and show that \texttt{PLUTO}/\texttt{IDEFIX} struggle to resolve higher-order modes of the buoyancy-related oscillations, weakening vortensity growth and the associated torque. We interpret this as a drawback of total-energy-conserving, finite-volume schemes. Our results indicate that a very high resolution or high-order scheme is required in shock-capturing codes in order to adequately capture this effect.

\end{abstract}



\section{Introduction}
\label{sec:introduction}

The observation of a pair of planets embedded in the circumstellar disk around PDS~70 \citep{keppler-etal-2018,haffert-etal-2019} has cemented the concept that planets form and grow in circumstellar disks. Their final radial location---once the disk has dispersed---is typically determined by their migration history through (and interaction with) the disk. This is especially important for low-mass planets, which can migrate rapidly \citep[e.g.,][]{ward-1997}. For a recent review see \cite{paardekooper-etal-2022}.

Planet--disk interaction can lead to several different sources of torques, which in turn drive the planet inwards or outwards \citep{ward-1997}. Typical examples are the Lindblad torques by planet-driven spiral wakes \citep{ward-1986,ogilvie-lubow-2002}, the horseshoe drag by the planet's corotating region \citep{ward-1991, masset-2001, paardekooper-papaloizou-2009} and the dynamical corotation torque experienced by a migrating planet in a disk with a radial vortensity gradient \citep{paardekooper-2014, mcnally-etal-2017}.

Different physical effects can give rise to additional sources of torques, such as thermal lobes in the planet's corotating region \citep{lega-etal-2014}, thermal input due to accretion heating \citep{benitez-etal-2015, masset-2017}, or overdensities around waves caused by the disk buoyancy response near the planet \citep{zhu-etal-2012}. The latter, which could potentially amount to a few tens of percent of the total torque, is the focus of this work.

In a recent study, \citet{mcnally-etal-2020} (hereafter \mcnally{}) showed that the torque associated with the disk buoyancy response can be significant, speeding up inward migration of low-mass planets. The torque was explained by identifying overdense lobes centered at a height $z\sim0.5\text{--}1\,H$, where $H$ is the disk scale height, near the planet's corotating region. Equivalently, the associated torque can be attributed to a vortensity enhancement in the planet's corotating region, weakening the stalling effect of the (outward) corotation torque \citep{mcnally-etal-2017}.

Modeling the buoyancy response of the disk can be challenging from a numerical point of view. As \mcnally{} showed, resolving the vortensity evolution, the associated dynamical corotation torque, and the planet's migration track requires high-resolution, three-dimensional global models of planet--disk interaction that are integrated for thousands of orbits. Such calculations are very computationally expensive, and especially so when additional physics such as radiation transport are included \citep{yun-etal-2022}. Therefore, great care should be exercised to ensure that the problem is sufficiently resolved by a given numerical scheme.

Both \mcnally{} and \citet{yun-etal-2022} modeled the effects of buoyancy-related torques on planet migration using \texttt{FARGO3D} \citep{benitez-llambay-etal-2016}, a code developed primarily to study highly supersonic shear flows such as protoplanetary disks. \texttt{FARGO3D} uses an operator-split upwind scheme on a staggered mesh, a method that has been successfully applied to planet--disk interaction for many years. Nevertheless, this is not the only approach that has been used to model planet--disk interaction: shock-capturing, Godunov-scheme codes such as \texttt{NIRVANA-III} \citep{ziegler-2004}, \texttt{PLUTO} \citep{mignone-etal-2007}, and \texttt{Athena++} \citep{stone-etal-2008} have been used extensively to study protoplanetary disks and planet--disk interaction as well. It is therefore worth investigating how different families of codes handle the subtle effects of the buoyancy response on planet migration.

In this study we compare the results of the global simulations of \mcnally{} to functionally identical models using \texttt{PLUTO}. Our primary aim is to identify the source of the buoyancy-related torque, and the reason the two codes give significantly different results. We complement these models with local, shearing-box simulations using \texttt{IDEFIX} \citep{lesur-etal-2023} and \texttt{FARGO3D}. Our goal there is to find the physical regime where these two codes recover similar solutions.

Our physical framework and numerical setup are described in Sect.~\ref{sec:physics-numerics}. We present our results in Sects.~\ref{sec:results-pluto-vs-fargo}~\&~\ref{sec:idefix-vs-fargo}, and discuss our findings in Sect.~\ref{sec:discussion}. We then provide a summary and our conclusions in Sect.~\ref{sec:summary}.

\section{Physics and numerics}
\label{sec:physics-numerics}

In this section we describe the physical framework and our numerical methods. Our setup closely replicates that of \mcnally{}, so we briefly summarize the relevant equations here. We refer the reader to \mcnally{} for a more detailed description. 

\subsection{Physics}
\label{sub:physics}

We consider a disk of ideal gas with adiabatic index $\gamma$ and mean molecular weight $\mu$ around a star with mass $\Mstar$. The gas has a volume density $\rho$, velocity field $\vel$ and pressure $P$ given by the ideal gas law $P = (\gamma-1)e$, where $e$ is the internal energy density. In a cylindrical coordinate system $\{R,\phi,z\}$, the inviscid Navier--Stokes equations then read
\begin{subequations}
	\label{eq:navier-stokes}
	\begin{align}
		\label{eq:navier-stokes-1}
		\DP{\rho}{t} + \vel\cdot\nabla\rho=-\rho\nabla\cdot\vel,
	\end{align}
	\begin{align}
	\label{eq:navier-stokes-2}
		\DP{\vel}{t}+ (\vel\cdot\nabla)\vel=-\frac{1}{\rho}\nabla P -\nabla(\Phi_\star+\Phi_\mathrm{p}),
	\end{align}
	\begin{align}
	\label{eq:navier-stokes-3}
		\DP{e}{t} + \vel\cdot\nabla e=-\gamma e\nabla\cdot\vel.
	\end{align}
\end{subequations}
Here, $\Phi_\star = -\G \Mstar/r$ is the potential of the star at spherical distance $r = \sqrt{R^2+z^2}$ from the star, $\Phi_\mathrm{p}$ is the potential of the planet, which we describe below, and $\G$ is the gravitational constant. We can now define the pressure scale height $H=\csiso/\OmegaK$, where $\csiso = \sqrt{P/\rho}$ is the isothermal sound speed and $\OmegaK=\sqrt{\G\Mstar/R^3}$ is the Keplerian frequency of the gas. The gas temperature is then $T = \mu\csiso^2/\Rgas$, where $\Rgas$ is the ideal gas constant.

Assuming $u_R=u_z=0$, a vertically isothermal profile ($\partial T/\partial z=0$), axisymmetry ($\partial/\partial\phi=0$), and radial power-law profiles for the midplane density $\rhomid$ and temperature such that
\begin{equation}
	\label{eq:rho-T}
	\rhomid(R) = \rho_0 \left(\frac{R}{R_0}\right)^p, \quad T(R) = T_0 \left(\frac{R}{R_0}\right)^q,
\end{equation}
we can derive an equilibrium state for the disk following \citet{nelson-etal-2013}
\begin{align}
	\label{eq:equilibrium}
	\rho^\text{eq}(R,z) &= \rhomid(R)\, \exp\left[-\frac{1}{h^2}\left(1-\frac{R}{r}\right)\right], \\
	u_\phi(R,z) &= R\OmegaK \left[1 + (p+q)h^2 + q\left(1-\frac{R}{r}\right)\right]^{1/2},
\end{align}
where $h=H/R$ is the disk aspect ratio. Finally, we define the surface density $\Sigma(R) = \int_{-\infty}^{\infty} \rho(R,z)\,\text{d}z$. 

We now embed a planet with mass $\Mp$ at distance $\Rp$ from the star. The planet is treated as a sphere with uniform density and radius $\epsilon=0.1H_\mathrm{p}$. We can then write the planetary potential at a distance $d=|\bm{r}-\bm{r}_\text{p}|$ from the planet using the prescription of \citet{klahr-kley-2006}, where a cubic spline is used for $d<\epsilon$ to prevent numerical singularities \citep[see also][]{yun-etal-2022}
\begin{equation}
	\label{eq:planet-potential}
	\Phi_\mathrm{p} = \begin{cases}
        -\G\Mp/d, & d \geq \epsilon \\
        -\G\Mp(3\epsilon^2-d^2)/(2d^3), & d < \epsilon.
    \end{cases}
\end{equation}

The indirect term by the planet--star system orbiting about their common center of mass is included. The planet is typically held on a fixed, circular orbit at $\Rp$, in which case the indirect term by the disk on the star is ignored. In models where the planet is allowed to migrate, that term is included \citep[see][]{crida-etal-2022}. Since we neglect the effect of self-gravity, we include the torque correction term of \citet{baruteau-masset-2008} in models with migrating planets.

In addition to global models, we study a local version of the problem using the well-known shearing box formalism \citep[see e.g.][]{hawley-etal-1995}. This formalism can be derived by considering a small box orbiting at the local Keplerian velocity at radius $r_0$, neglecting any curvature effects (hence working in Cartesian coordinates) and expanding the stellar potential in a small parameter $L/r_0$, where $L$ is the size of the box. The local version of Eqs.~\eqref{eq:navier-stokes} then reads:
\begin{subequations}
	\label{eq:navier-stokes-local}
	\begin{align}
		\label{eq:navier-stokes-local-1}
		\DP{\rho}{t} + \vel\cdot\nabla\rho=-\rho\nabla\cdot\vel,
	\end{align}
	\begin{align}
	\label{eq:navier-stokes-local-2}
		\DP{\vel}{t}+ (\vel\cdot\nabla)\vel=-2\bm{\Omega}_0\times \vel -\frac{1}{\rho}\nabla P -\nabla(\Phi_0+\Phi_\mathrm{p}) + \nu \nabla^2\vel,
	\end{align}
	\begin{align}
	\label{eq:navier-stokes-local-3}
		\DP{e}{t} + \vel\cdot\nabla e=-\gamma e\nabla\cdot\vel +\chi\nabla\cdot\left[\rho\nabla\left(\frac{e}{\rho}\right)\right],
	\end{align}
\end{subequations}
where $\bm{\Omega}_0$ is the angular velocity of the shearing box and $\Phi_0 = -3\Omega_0^2x^2/2 + \Omega_0^2 z^2/2$ is the tidal potential. We have also included a kinematic viscosity $\nu$ and thermal diffusivity $\chi$. We use the same planet potential as in the global case (see Eq.~\eqref{eq:planet-potential}), with the planet located at the origin of our coordinate frame. A vertically isothermal equilibrium solution is given by
\begin{equation}
	\label{eq:local-density}
    \rho(z) = \rho_0\exp(-\Omega_0^2z^2/(2\csiso^2)),
\end{equation}
together with $P = \csiso^2\rho$ and $\vel = (0, -3\Omega_0 x/2,0)^T$. The isothermal sound speed $\csiso$ is initially constant throughout the box. 

\subsection{Dynamical corotation torque}
\label{sub:corotation-torque}

\citet{mcnally-etal-2017} have shown that a migrating planet experiences a torque by the material inside the corotating region given by
\begin{equation}
	\label{eq:corotation-torque}
	\Gamma_h  = 2\pi\left(1-\frac{\varpi(\Rp)}{\varpi_h}\right)\Sigma_\mathrm{p}\Rp^2 \xh\Omega_\mathrm{p}\left(\D{\Rp}{t} - u_R\right),
\end{equation}
where $\xh$ is the horseshoe half-width \citep{paardekooper-etal-2010} given by
\begin{equation}
	\label{eq:horseshoe-width}
	\xh = \frac{1.1}{\gamma^{1/4}}\left(\frac{0.4}{\epsilon/H_\text{p}}\right)^{1/4}\sqrt{\frac{\Mp}{\hp\Mstar}}\Rp,\qquad \epsilon = 0.6H_\text{p}.
\end{equation}
Here, $\epsilon$ is chosen to match 3D models \citep{mueller-kley-2012}, $\varpi(\Rp)$ is the vortensity at $\Rp$ for an unperturbed disk, and $\varpi_\mathrm{h}$ is the characteristic vortensity enclosed in the planet's corotating region. For a vertically integrated disk the vortensity is given by $\varpi=(\nabla\times \vel)/\Sigma\cdot\hat{z}$, while for a 3D disk the equivalent quantity was shown by \citet{masset-llambay-2016} to be
\begin{equation}
	\label{eq:vortensity-3d}
	\varpi = \left[\int\limits_{-\infty}^\infty\frac{\rho}{(\nabla\times\vel)\cdot\hat{z}}\mathrm{d}z\right]^{-1}.
\end{equation}
It can then be shown that for both 2D and 3D, the unperturbed vortensity is approximately $\varpi_0(R)\approx\frac{1}{2}\OmegaK/\Sigma^\text{eq}$, with $\Sigma^\text{eq} = \int_{-\infty}^{\infty}\rho^\text{eq}\text{d}z$.

\subsection{Numerics}
\label{sub:numerics}

We mainly use the \texttt{PLUTO v4.4} code \citep{mignone-etal-2007}, with comparison models using \texttt{FARGO3D} \citep{benitez-llambay-etal-2016} and \texttt{IDEFIX} \citep{lesur-etal-2023}. \texttt{PLUTO} and \texttt{IDEFIX} use a conservative, finite-volume, shock-capturing approach, by solving the Riemann problem across all cell interfaces. By contrast, \texttt{FARGO3D} uses a finite-difference method with a staggered mesh to achieve momentum conservation to machine precision. All codes employ the FARGO method by \citet{masset-2000}, implemented in \texttt{PLUTO} by \citet{mignone-etal-2012}. This enables considerable speedup by subtracting the background disk rotation before solving Eqs.~\eqref{eq:navier-stokes}, relaxing the strict timestep limitation by the quickly-rotating inner boundary. We note that in the \texttt{PLUTO} and \texttt{IDEFIX} implementations, the conservative form of Eqs.~\eqref{eq:navier-stokes} is solved instead \citep[see e.g., Eqs.~(1)~\&~(24) in][]{mignone-etal-2007}.

The version of \texttt{FARGO3D} that we use includes the specific entropy formulation described in \citet{mcnally-etal-2019b} (see Appendix~A therein), which better conserves entropy in the absence of shocks. In \texttt{PLUTO}, we use the \texttt{ENTROPY\_SWITCH} option to achieve the same effect. By default, we use second-order accurate time-marching (\texttt{RK2}) and reconstruction (\texttt{LINEAR}) schemes in \texttt{PLUTO}, but we investigate several numerical combinations in our study (see Sect.~\ref{sub:different-numerics}). Both codes use the flux limiter by \citet{vanleer-1974}.

We use \texttt{PLUTO} to run global simulations identical to the \texttt{FARGO3D} setup of \mcnally{} to facilitate a fair comparison between the two codes. In these models, our grid and physical setup is identical to that of \mcnally{}, with a grid spanning $r\in[0.8,4.0]$~au, $\theta\in[\pi/2-\arctan(5h), \pi/2]$, and $\phi\in[0,2\pi]$ with $N_r\times N_\theta\times N_\phi = 805\times125\times3141$ for most models, achieving a resolution of 25 cells per scale height. We assume a constant aspect ratio $h=0.05$ (i.e., $q=-1$) and a density profile that corresponds to $\rho_0=0.003\,\Msun/\text{au}^3$ and $p=-3/2$, or $\Sigma = 3340\,\text{g}/\text{cm}^2
\,(R/\text{au})^{-1/2}$, with $R_0=1$\,au. The planet is held on a fixed circular orbit at $\Rp=2$~au for at least 200 orbits before it is allowed to migrate. Finally, we use wave-damping boundary zones at the top and radial sides of the domain similar to \citet{mcnally-etal-2019a}. In \texttt{FARGO3D} this is implemented as a linear viscous pressure \citep{stone-norman-1992}, while in \texttt{PLUTO} we use the wave-killing method described in \citet{devalborro-etal-2006} with a damping timescale of $0.1$ local orbits.

In addition to our global studies, we employ the shearing box formalism to compare \texttt{IDEFIX} to \texttt{FARGO3D}. The unit of distance is taken to be $H_0=\csiso(t=0)/\Omega_0$, and the domain considered is $x \in (0, 10)\,H_0$, $y\in (-20,20)\,H_0$, and $z \in (0, 3.873)\,H_0$ following \citet{zhu-etal-2012}. This domain is covered at a minimum of $N_x\times N_y\times N_z=160\times640\times80$ grid cells, yielding a resolution of 16 cells per $H_0$. Boundaries are taken to be periodic in $y$, and in $z$ we employ outflow boundaries at the top of the domain, while a symmetry condition is forced at $z=0$. Towards the outer boundary in $x$, over the last $2H_0$ we employ wave damping as described in \cite{devalborro-etal-2006}. At $x=0$, we force the solution to be point-symmetric around the origin \citep[see][]{zhu-etal-2012}

Our local models use the vanilla version of \texttt{FARGO3D}, without the specific entropy formulation. For \texttt{IDEFIX}, we use standard linear reconstruction with \texttt{RK2} time-marching. In the case of non-zero diffusion coeffisions, we use super-timestepping \citep{alexiades-etal-1996} to alleviate constraints on the timestep from viscosity and thermal diffusion. We do not employ the FARGO method in \texttt{IDEFIX}. While this leads to smaller time steps, we have checked that \texttt{FARGO3D} with and without the FARGO method gives identical results, which indicates that numerical diffusion associated with time stepping is not a major issue in this problem.

\section{Global models: \texttt{PLUTO} vs.~\texttt{FARGO3D}}
\label{sec:results-pluto-vs-fargo}

In this section we present a comparison between our global models with \texttt{PLUTO} and the models by \mcnally{} using \texttt{FARGO3D}. We begin with a comparison of our fiducial models, and then investigate the effect of different numerical schemes in \texttt{PLUTO}. 

\subsection{Fiducial models}
\label{sub:pluto-vs-fargo-fiducial}

In their study, \mcnally{} demonstrated that the buoyancy response by the disk to the embedded planet drives a time-dependent vortensity enhancement in the planet's corotating region, resulting in an excess negative corotation torque and the speeding up of inward migration. To measure the torque acting on the planet we sum all contributions from the disk inside the active domain
\begin{equation}
	\label{eq:torque}
	\Gamma = \G\Mp \int\limits_{\text{disk}} \frac{\bm{r}\times\bm{r}_\text{p}}{d^3}\rho\text{d}V, \qquad \bm{d} = \bm{r}-\bm{r}_\text{p},
\end{equation}
and normalize the result by a reference value $\Gamma_0$ \citep[e.g.,][]{kley-nelson-2012}
\begin{equation}
	\label{eq:torque0}
	\Gamma_0 = \left(\frac{\Mp}{\hp\Mstar}\right)^2 \Sigma_\text{p} \Rp^4 \Omega_\text{p}^2.
\end{equation}

\begin{figure*}
	\includegraphics[width=\textwidth]{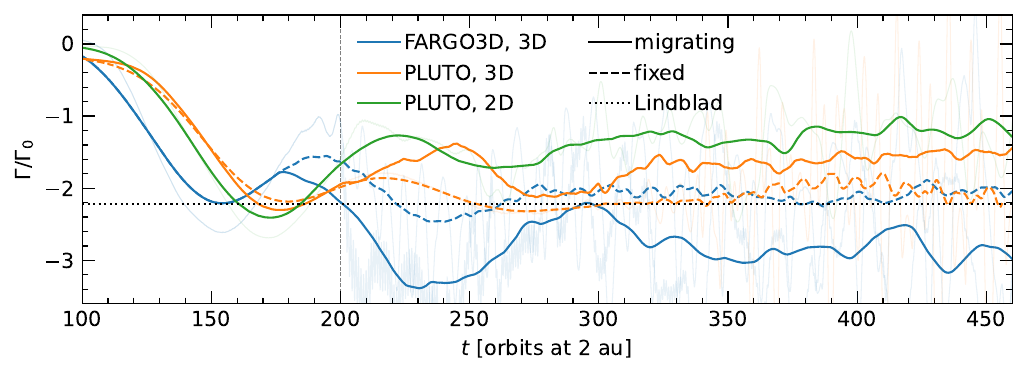}
	\caption{Torques acting on planets in our fiducial 3D models for \texttt{FARGO3D} (blue) and \texttt{PLUTO} (orange), for fixed (dashed) and migrating (solid) planets. The two codes agree very well for the fixed case, but differ substantially when the planet is allowed to migrate. The torque acting on the migrating planet in the 3D \texttt{FARGO} model is stronger than the fixed case, indicating feedback by the disk buoyancy response \mcnallyp{}.
    For \texttt{PLUTO}, the 3D model shows a slightly stronger torque than a corresponding 2D setup (green), implying that this effect is present but not well resolved. The dots follow Eq.~(14) in \citet{paardekooper-etal-2010}.}
	\label{fig:fiducial-torques}
\end{figure*}

Figure~\ref{fig:fiducial-torques} shows a comparison between the two codes for models where the planet is either fixed or allowed to migrate. We find that the two codes agree very well for the fixed planet case in 3D (blue and orange dashed curves), but show significant differences when the planet is allowed to migrate (solid curves). In particular, the \texttt{PLUTO} model (orange) shows a much weaker torque than the \texttt{FARGO3D} model (blue). 
Nevertheless, the 3D \texttt{PLUTO} model shows a more negative torque than a corresponding 2D, vertically integrated model (shown in green), implying that the torque-generating mechanism found in the 3D models of \mcnally{} is operating to an extent. We note that the 2D model has been run with \texttt{PLUTO}, but the torque and migration tracks agree very well with a 2D model using \texttt{FARGO3D}.

\begin{figure}
	\includegraphics[width=\columnwidth]{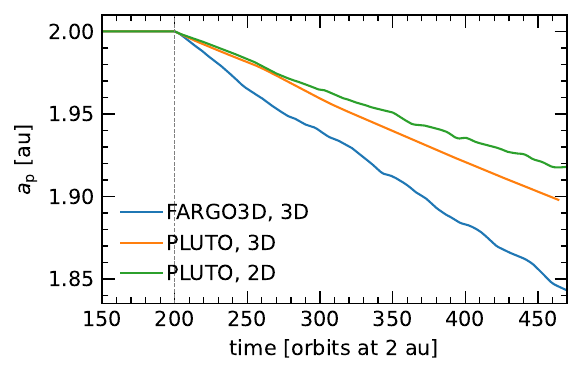}
	\caption{Migration tracks for the different models shown in Fig.~\ref{fig:fiducial-torques}. In the 3D model with \texttt{FARGO3D} (blue) the planet migrates inwards faster than a 2D model (green) due to the stronger negative corotation torque acting on it. Migration is much slower for the 3D \texttt{PLUTO} model (orange), suggesting that the effect that speeds up the planet in \texttt{FARGO3D} is not as strong in \texttt{PLUTO}.}
	\label{fig:fiducial-migration}
\end{figure}

In Fig.~\ref{fig:fiducial-migration} we compare the migration tracks for all models shown in Fig.~\ref{fig:fiducial-torques}. Our results remain consistent: the planet in the 3D \texttt{PLUTO} model migrates slower than that in the \texttt{FARGO3D} model, and slightly faster than the corresponding 2D model.
We can therefore conclude that the vortensity generation by the disk buoyancy response is present, but too weak to speed up inward migration in our \texttt{PLUTO} models at the same rate as in \texttt{FARGO3D}.

\begin{figure}
	\includegraphics[width=\columnwidth]{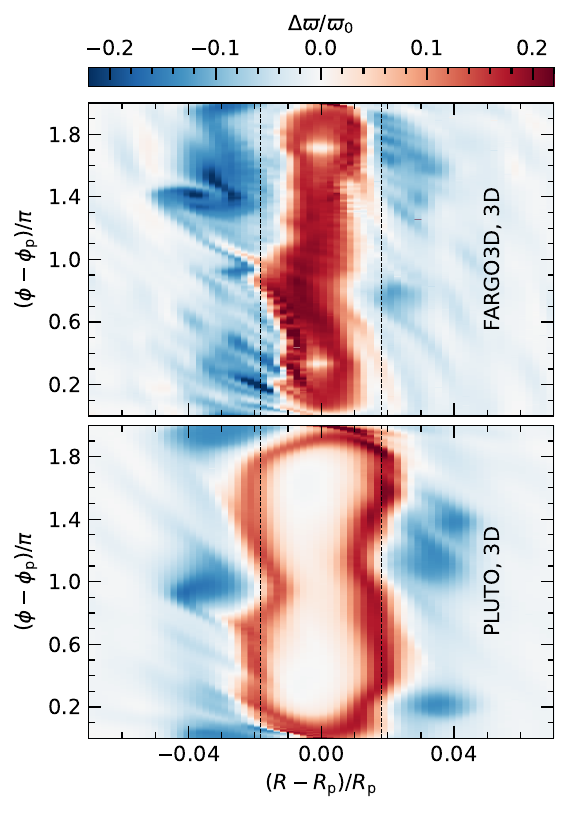}
	\caption{Two-dimensional heatmaps of the vortensity deviations in the planet's horseshoe region for the fixed 3D models at $t=500\,\Pp$. Excess vortensity is concentrated about $\Rp$ for \texttt{FARGO3D}, but instead follows streamlines at the edge of the corotating region for \texttt{PLUTO}.}
	\label{fig:fiducial-vortensity}
\end{figure}

\begin{figure}
	\includegraphics[width=\columnwidth]{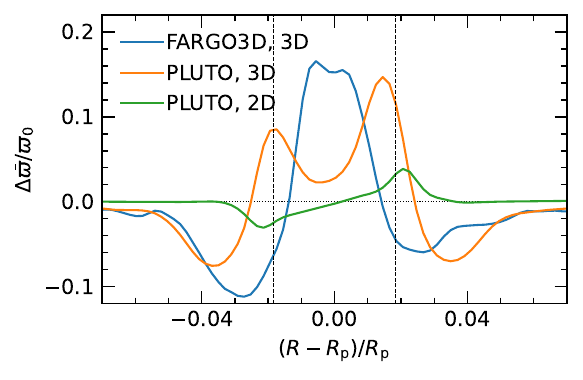}
	\caption{Azimuthally averaged vortensity deviations in the planet's corotating region for fixed planets at $t=500\,\Pp$. Even though there is a net $\varpi$ growth in 3D models, their structure is quite different. Vertical dots mark $\pm\xh$.}
	\label{fig:fiducial-vortensity-mean}
\end{figure}

Given that our results agree very well between codes for the fixed planet case (as well as in 2D, not shown here), we expect that the two codes resolve the buoyancy response to a different extent, and therefore the vortensity growth rate is different. Figure~\ref{fig:fiducial-vortensity} shows the vortensity structure in the corotating region of the planet for the \texttt{FARGO3D} (top) and \texttt{PLUTO} (bottom) models, showing they are very different between the two codes. Specifically, the vortensity enhancement in \texttt{FARGO3D} is concentrated at $\Rp$, while in the case of \texttt{PLUTO} we see closed streamlines with excess vortensity at the edges of the corotating region. Figure~\ref{fig:fiducial-vortensity-mean} shows a radial profile of the azimuthally averaged vortensity in the corotating region to better highlight the magnitude of the vortensity excess.

\begin{figure}
	\includegraphics[width=\columnwidth]{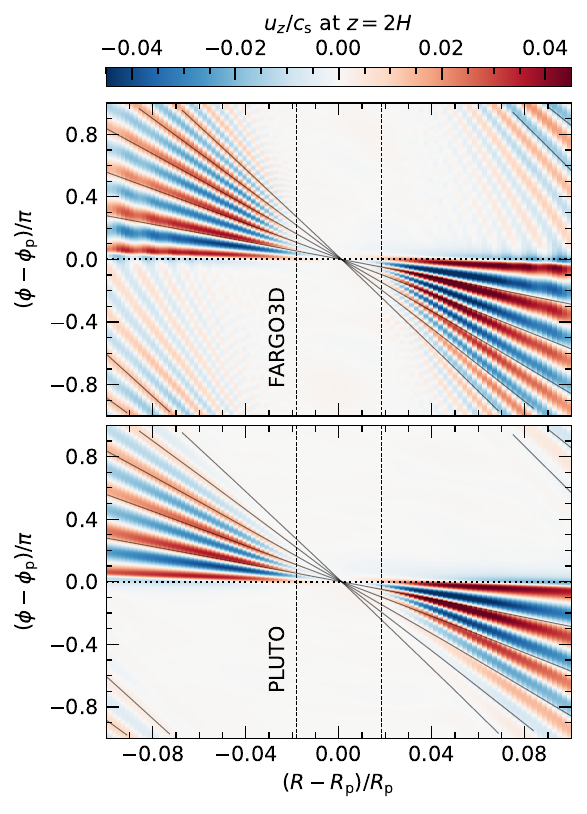}
	\caption{Two-dimensional heatmaps of the vertical gas velocity component $u_z$ at a height $z=2H$ for both codes, showing oscillations due to the disk buoyancy response. Diagonal black lines mark wavefronts of constant phase (see Eq.~\eqref{eq:buoyancy-phi}), and vertical black lines denote the limits of the corotating region $\pm \xh$. \texttt{FARGO3D} shows more buoyancy modes than \texttt{PLUTO}.}
	\label{fig:fiducial-uz}
\end{figure}

\begin{figure}
	\includegraphics[width=\columnwidth]{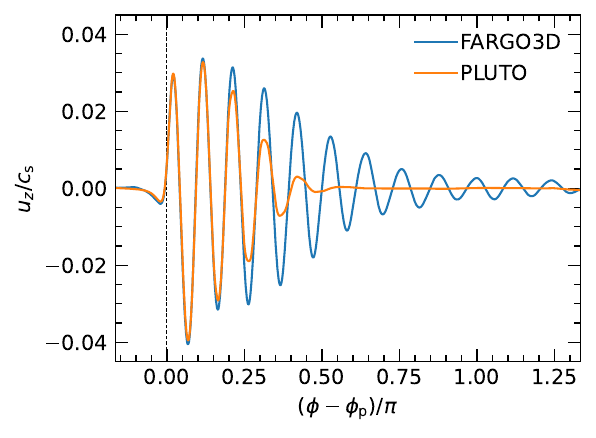}
	\caption{Gas $u_z$ at $R=\Rp-2\xh$ and $z=2H$, normalized to the local sound speed. Similar to Fig.~\ref{fig:fiducial-uz}, we see more oscillations in \texttt{FARGO3D}. The planet is marked by a vertical black line, and the undulations about it are due to the planet's spiral arm.}
	\label{fig:fiducial-uz-mean}
\end{figure}

To understand why the vortensity structure in the corotating region is so different, we turn to investigating the buoyancy response of the disk to the planet. As material approaches the planet along the azimuthal direction at a height $z_0$ it receives a ``kick'' due to the planet's potential, and therefore oscillates about $z_0$ with a frequency given by the Brunt--V{\"a}is{\"a}l{\"a} frequency
\begin{equation}
	\label{eq:buoyancy-frequency}
	N(z) = \sqrt{\frac{g_z}{\gamma}\frac{\partial}{\partial z}\left[\ln
	\left(\frac{P}{\rho^\gamma}\right)\right]}, \qquad g_z = -\frac{\partial\Phi_\star}{\partial z},
\end{equation}
with resonant modes being excited on lines of constant phase at azimuth $\phi_b$ given by
\begin{equation}
	\label{eq:buoyancy-phi}
	\phi_b(R,z) = \pm 2n\pi\sqrt{\frac{\gamma}{\gamma-1}} \frac{\Omega_\mathrm{p}-\Omega}{\OmegaK}\frac{H}{z}\left(1+\frac{z^2}{R^2}\right)^{3/2},
\end{equation}
where $n$ is an integer \citep[see also][]{zhu-etal-2015}. We now plot the vertical velocity component of the gas $u_z(R,\phi)$ at a height $z=2H$ after 200 planetary orbits for both codes and show the comparison in Fig.~\ref{fig:fiducial-uz}. We find that there are fewer modes visible for \texttt{PLUTO}, while \texttt{FARGO3D} shows more (albeit noisier) structure. We illustrate this more quantitatively in Fig.~\ref{fig:fiducial-uz-mean}, where we plot $u_z(\phi)$ at $R=\Rp-2\xh\approx0.96\,\Rp$ and $z=2H$, showing that there are fewer and weaker oscillations resolved by \texttt{PLUTO}. This behavior is consistent regardless of $R$ and $z$, and most likely relates to the total variation diminishing (TVD) nature of Godunov schemes, which might damp wave-like behavior overzealously to avoid spurious oscillations at the grid scale. By contrast, the algorithm used by \texttt{FARGO3D} results in possibly more accurate but also noisier structures.

\begin{figure}
	\includegraphics[width=\columnwidth]{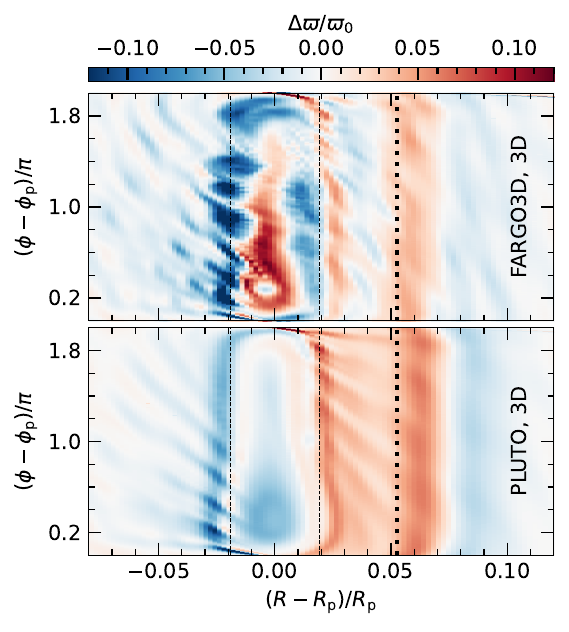}
	\caption{Similar to Fig.~\ref{fig:fiducial-vortensity}, for models with migrating planets at $R=1.9$\,au. A vortensity excess is clearly visible for \texttt{FARGO3D}. The dotted vertical line marks the planet's initial location.}
	\label{fig:fiducial-vortensity-mobile}
\end{figure}

Assuming that these buoyancy modes are responsible for the vortensity growth in the corotating region, it becomes clear that the amplitude and number of modes resolved will decide the rate of vortensity growth as well as its radial structure. Given that higher-order modes $n\gg1$ oscillate more tightly around the planet, they have the potential to both generate vortensity more efficiently inside the corotating region as well as closer to $\Rp$. Since \texttt{PLUTO} resolves fewer modes, the vortensity growth is slower and more spread out in radius, resulting in a smaller excess that is confined to the edges of the corotating region. By contrast, in \texttt{FARGO3D} the vortensity excess is more concentrated at $\Rp$. This can also be inferred from Fig.~\ref{fig:fiducial-vortensity-mean}. As the planet migrates and the corotating region shrinks to a tadpole shape \citep{masset-papaloizou-2003, papaloizou-etal-2007}, the excess trapped in the corotating region is higher in \texttt{FARGO3D} than in \texttt{PLUTO} (see Fig.~\ref{fig:fiducial-vortensity-mobile}), resulting in a stronger negative torque (as shown in Fig.~\ref{fig:fiducial-torques}).

\subsection{Vortensity forcing}

We now investigate whether the vortensity growth is tied to the buoyancy response. We start with the vortensity equation, obtained by taking the curl of the momentum equation \eqref{eq:navier-stokes-2} and dividing by $\rho$
\begin{equation}
	\label{eq:vortensity-equation}
	\D{\bm{\omega}_\rho}{t} = \left(\bm{\omega}_\rho\cdot\nabla\right)\vel + \frac{\nabla\rho\times\nabla P}{\rho^3}, \qquad \bm{\omega}_\rho = \frac{\nabla\times\vel}{\rho},
\end{equation}
where $\omega_\rho = \bm{\omega}_\rho\cdot\hat{z}$ is the potential vorticity and $\text{d}/\text{d}t$ is the material derivative. The first and second term on the right hand side relate to vortex stretching and baroclinic forcing, respectively. Given the vertically slanted nature of the buoyancy responce \citep{zhu-etal-2012} and the presence of a global pressure gradient, both terms are expected to correlate with the buoyancy response and possibly contribute to vortensity growth.

\begin{figure}
	\includegraphics[width=\columnwidth]{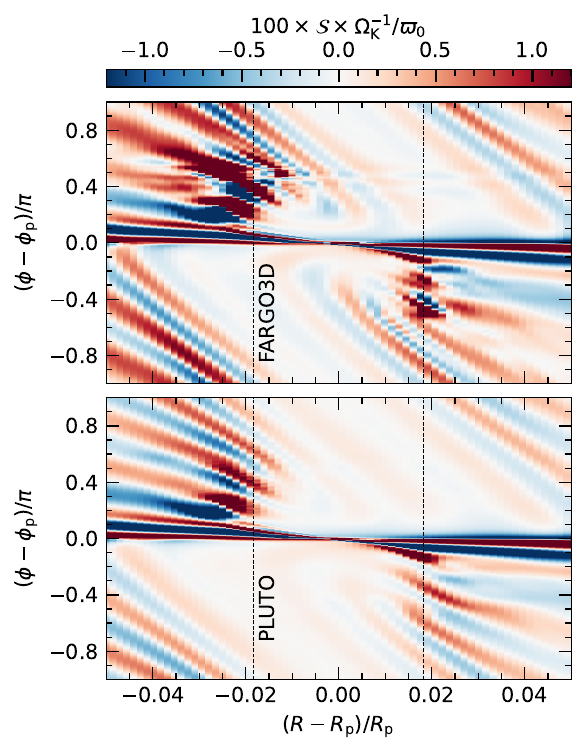}
	\caption{The vortensity source term through Eq.~\eqref{eq:vortensity-source}, normalized to a dimensionless quantity. We note the more saturated colors on the top panel (\texttt{FARGO3D}), indicating stronger vortensity generation. A horizontal ``glitch'' at $\phi/\pi\pm 0.4$ is the result of spiral arms by emerging vortices.}
	\label{fig:source-term}
\end{figure}

Using Eq.~\eqref{eq:vortensity-3d}, we can then approximate the vortensity growth rate as
\begin{equation}
	\label{eq:vortensity-source}
	\D{\varpi}{t} \approx \varpi^2\int\limits_{-\infty}^\infty \frac{1}{\omega_\rho^2}\,\left[\left(\bm{\omega}_\rho\cdot\nabla\right)\vel + \frac{\nabla\rho\times\nabla P}{\rho^3}\right]\cdot\hat{z}\,\mathrm{d}z = \mathcal{S}(R,\phi).
\end{equation}
We then plot this quantity $\mathcal{S}$ for both codes in Fig.~\ref{fig:source-term}, using high-resolution models from Sect.~\ref{sub:different-numerics} for clearer results (the numerical and physical setups are otherwise exactly the same). We find that vortensity is generated along the planet-driven spiral shocks, which is to be expected, as well as along diagonals about the planet that coincide with the buoyancy modes in Eq.~\eqref{eq:buoyancy-phi}. This confirms that the vortensity generation is due to both ``vortex stretching'' and baroclinic forcing by the disk's buoyancy response inside the planet's corotating region, with baroclinic forcing contributing to roughly 4--5\% of the total vortensity growth. We find stronger but significantly noisier vortensity generation for \texttt{FARGO3D} and larger $n$, consistent with our expectations above, supporting the idea that buoyancy modes might be overdamped in \texttt{PLUTO} (but also underdamped in \texttt{FARGO3D}). The source term also reaches closer to $\Rp$ for \texttt{FARGO3D}, which is consistent with the more concentrated vortensity excess in Fig.~\ref{fig:fiducial-vortensity-mean}.

Based on our findings, we expect that the primary reasons for the two codes to differ are of numerical nature. However, it is still not clear which of the two codes offers the ``more correct'' answer, as \texttt{PLUTO} tends to overdamp oscillations for this problem (and vice versa for \texttt{FARGO3D}). Since \texttt{PLUTO} offers a wide range of numerical options, we experiment with various configurations in the next section.

\subsection{Different numerics}
\label{sub:different-numerics}

In the previous section we established that the differences between \texttt{PLUTO} and \texttt{FARGO3D} are primarily not physical, but instead most likely owe to the very different numerical methods used in each code. To investigate further, we run a series of tests between the two codes.

We begin with a resolution analysis, to test whether \texttt{PLUTO} recovers the behavior in \texttt{FARGO3D} by simply increasing the resolution. For these tests we let the planet grow over one orbit and integrate for 10 planetary orbits. This allows enough time for the buoyancy patterns to be established in $u_z$, which we plot at $t=10\,P_\text{p}$.

\begin{figure}
	\includegraphics[width=\columnwidth]{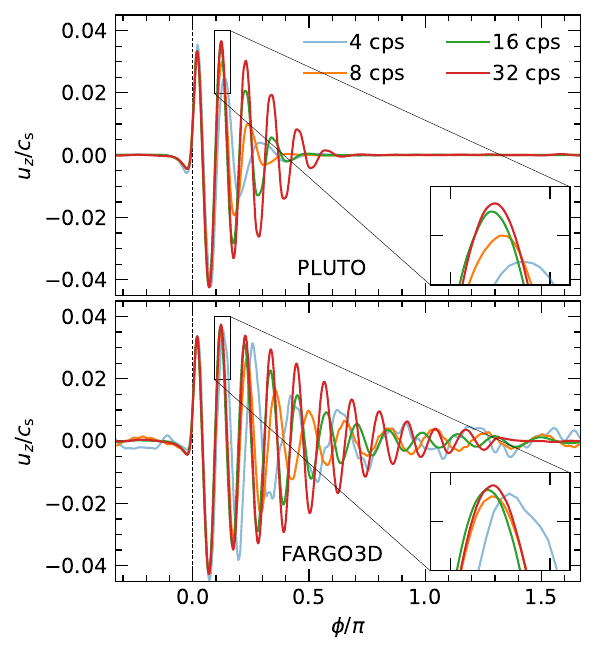}
	\caption{Gas $u_z$ similar to Fig.~\ref{fig:fiducial-uz-mean} for different resolutions in units of cells per scale height (cps). Both codes resolve the $n=1$ mode well at 16~cps (see insets), with \texttt{FARGO3D} showing the same amplitude for all resolutions (albeit a different phase for 4~cps). \texttt{FARGO3D} consistently resolves more modes than \texttt{PLUTO}, but also shows spurious oscillations at low resolutions (4--8~cps).}
	\label{fig:uz-resolutions}
\end{figure}

Figure~\ref{fig:uz-resolutions} shows $u_z(\phi)$ at $R=\Rp-2\xh$ and $z=2H$, similar to Fig.~\ref{fig:fiducial-uz-mean}. We find that the $n=1$ mode is already resolved at low resolution for both codes, and they both resolve more buoyancy modes as the resolution increases. However, we also find that \texttt{FARGO3D} consistently resolves more and stronger modes at a given resolution compared to \texttt{PLUTO}, at the cost of some spurious oscillations far from the planet that become weaker with increasing resolution. In addition, \texttt{FARGO3D} achieves convergence for a given mode at a lower resolution than is necessary for \texttt{PLUTO}. Finally, we find that the phase of the buoyancy response is better captured at higher resolution, with modes packing together more tightly in both codes. Given that numerical noise in \texttt{FARGO3D} gives way to physical undulations at higher resolution, and the amplitude of the latter converges faster with increasing cell count in \texttt{FARGO3D}, we conclude that \texttt{FARGO3D} will resolve more and stronger buoyancy modes than \texttt{PLUTO} at any given resolution.

\begin{figure}
	\includegraphics[width=.99\columnwidth]{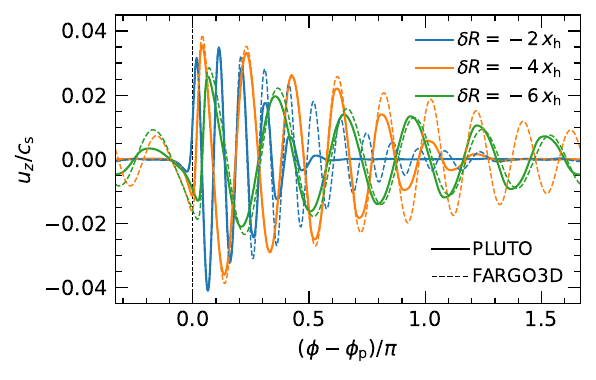}
	\caption{Gas $u_z$ similar to Fig.~\ref{fig:uz-resolutions}, at different distances from the planet $\delta R = R-\Rp$. \texttt{FARGO3D} once again shows stronger oscillations regardless of $\delta R$.}
	\label{fig:uz-radii}
\end{figure}

In Fig.~\ref{fig:uz-radii} we then compare $u_z$ at different radii or heights for both codes at a resolution of 16 cells per scale height. We find that both codes agree very well around the planetary spirals and $n=1$ modes but are very different when the focus is buoyancy modes with $n>1$, especially closer to $\Rp$.

\begin{figure}
	\includegraphics[width=\columnwidth]{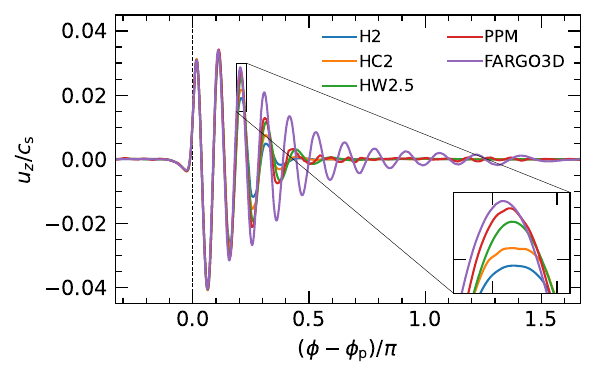}
	\caption{Gas $u_z$ similar to Fig.~\ref{fig:uz-resolutions} for different solvers in \texttt{PLUTO}. The 3rd-order \texttt{PPM} setup achieves best results, but still not as good as \texttt{FARGO3D}.}
	\label{fig:uz-numerics}
\end{figure}

This behavior implies that \texttt{FARGO3D} indeed recovers a more accurate solution for this problem than \texttt{PLUTO}, and especially so at higher resolution where the numerical shortcomings of \texttt{FARGO3D} reduce to the grid scale. With that in mind, we shift to \texttt{PLUTO} and run several tests with different numerical configurations at a resolution of 16 cells per scale height ($N_r\times N_\theta\times N_\phi = 528\times80\times2048$). We choose this lower resolution to significantly reduce computational costs, given that the results are qualitatively similar regardless of resolution.

From less to more accurate and computationally expensive, the numerical setups we used in \texttt{PLUTO} are summarized in Table~\ref{table:setups}. We note that not all models shown in this table are shown in plots. In that case, this implies that such a setup offers negligible gain compared to the setup right below it in the table.

\begin{table}
	\begin{center}	
		\caption{Configuration flags used in \texttt{PLUTO} to test the effects of numerics, sorted by ascending accuracy. The \texttt{H2} setup is our default, and is used for every run unless otherwise stated.}
		\label{table:setups}
		\begin{tabular}{c | c | c | c | c}
			\hline
			tag &  reconstruction & timestepping & solver & \texttt{misc.} \\\hline
			\texttt{H2} & \texttt{LINEAR} & \texttt{RK2} & \texttt{hllc} & (default)\\ 
			\texttt{HC2} & \texttt{LINEAR} & \texttt{RK2} & \texttt{hllc} & \texttt{CHAR\_LIMITING}\\ 
			\texttt{HL2.5} & \texttt{Lim03} & \texttt{RK2} & \texttt{hllc} & \texttt{CHAR\_LIMITING}\\ 
			\texttt{RW2.5} & \texttt{WENO3} & \texttt{RK2} & \texttt{roe} & \texttt{CHAR\_LIMITING}\\ 
			\texttt{RW3} & \texttt{WENO3} & \texttt{RK3} & \texttt{roe} & \texttt{CHAR\_LIMITING}\\ 
			\texttt{HW2.5} & \texttt{WENO3} & \texttt{RK2} & \texttt{hllc} & \texttt{CHAR\_LIMITING}\\ 
			\texttt{PPM} & \texttt{PARABOLIC} & \texttt{RK3} & \texttt{hllc} & \texttt{CHAR\_LIMITING}\\ 
		\end{tabular}
	\end{center}
\end{table}

Similar to above, we plot $u_z(\phi)$ at $R=\Rp-2\xh$ and $z=2H$ in Fig.~\ref{fig:uz-numerics}. All \texttt{PLUTO} models are compared against the corresponding \texttt{FARGO3D} model, showing that more accurate numerical setups result in more and stronger buoyancy modes. This supports the idea that \texttt{FARGO3D} recovers a more accurate solution, if one is willing to look past the numerical oscillations inherent in that code.

Interestingly, models \texttt{HL2.5}, \texttt{RW2.5}, \texttt{RW3}, and \texttt{HW2.5} showed nearly identical results, highlighting the importance of a higher-order reconstruction rather than timestepping. However, it is unexpected that the \texttt{HW2.5} model offered slightly better results than the others listed in this paragraph, resolving each mode with an amplitude $\approx0.7\%$ larger even though model \texttt{RW3} should be more accurate due to a less diffusive solver (\texttt{roe} instead of \texttt{hllc}) and a higher-order timestepping scheme (\texttt{RK3} instead of \texttt{RK2}). We did not investigate further on why that was the case.

\subsection{Section summary}

A comparison between \texttt{PLUTO} and \texttt{FARGO3D} showed that \texttt{PLUTO} results in a smooth, linear flow around the planet but also does not resolve or incorrectly damps oscillations due to the disk buoyancy response by the planet. On the contrary, these modes are excited more accurately in \texttt{FARGO3D}, and their vortensity generating propensity is more correctly captured, at the cost of numerical noise being present. These patterns are consistent for very low to high resolution, with \texttt{FARGO3D} consistently recovering more and stronger buoyancy modes.

Using different numerical configurations in \texttt{PLUTO}, we showed that the reason for the differences between the two codes is tied to their numerical approach, which is fundamentally different for each code (shock-capturing, finite-volume Godunov schemes in \texttt{PLUTO} versus staggered mesh, finite-difference upwind methods in \texttt{FARGO3D}). \texttt{PLUTO} recovers more modes with higher-order schemes or more accurate solvers, but \texttt{FARGO3D} nevertheless achieves convergence for a given buoyancy mode at typically lower resolutions. 

\FloatBarrier

\section{Shearing box models: \texttt{IDEFIX} vs.~\texttt{FARGO3D}}
\label{sec:idefix-vs-fargo}

In this section we describe the results using our local setup. We focus on a planet mass $\Mp$ such that $\G\Mp = 0.0058~\Omega_0^2H_0^3$, following \cite{zhu-etal-2012}. The planet radius is set to $\epsilon=0.7\,H_0$, to make sure this length scale is resolved for all resolutions studied. Our standard resolution has an approximate cell size of $\Delta x = H_0/16$, so that even when we go down a factor of 2 in resolution we still resolve the planet potential by several grid cells. 

\begin{figure}
	\includegraphics[width=\columnwidth]{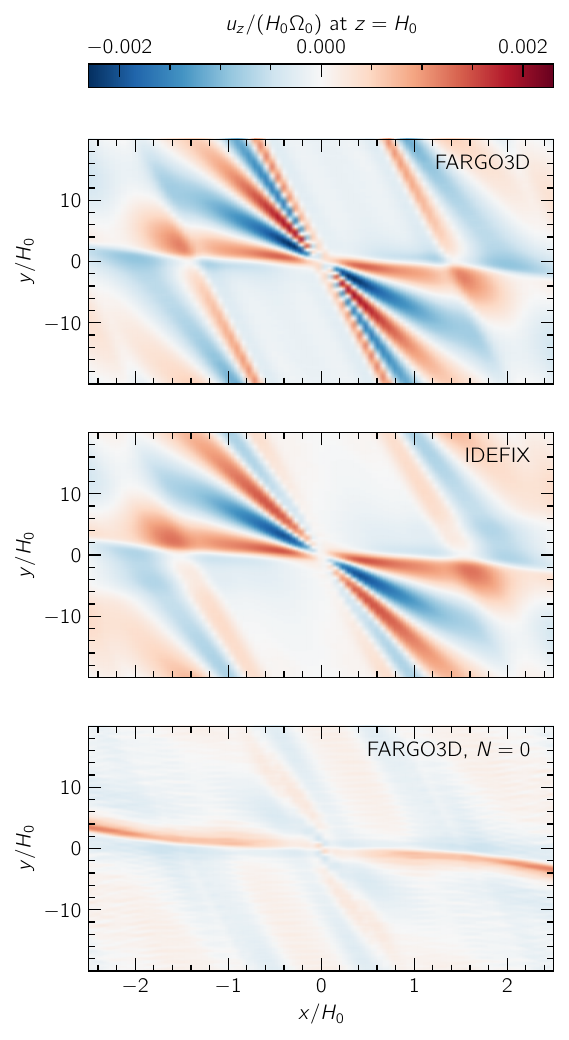}
	\caption{Two-dimensional heatmaps of the vertical gas velocity component $u_z$ at a height $z=H_0$ at $\Omega_0t=20$, using \texttt{FARGO3D} (top panel), \texttt{IDEFIX} (middle panel) and \texttt{FARGO3D} with a polytropic initial condition (bottom panel). The top two panels show oscillations due to the disk buoyancy response.}
	\label{fig:local_vz_2d}
\end{figure}

\subsection{The case without explicit diffusion}

In Fig.~\ref{fig:local_vz_2d} we show the vertical component of the gas velocity at $\Omega_0 t=20$ for our standard resolution of $\Delta x = H_0/16$, with no diffusion ($\nu=\chi=0$). This time was chosen such that a well-defined buoyancy response has developed, but the distortions due to the periodic $y$ boundaries remain limited. The top two panels show results obtained with an isothermal initial disc profile, using \texttt{FARGO3D} (top panel) and \texttt{IDEFIX} (middle panel). In the case of \texttt{IDEFIX}, the operator splitting technique used to integrate the vertical component of the stellar gravity makes it hard to start from exact numerical hydrostatic equilibrium \citep[this is a well-known problem, see e.g.][]{bale-etal-2003, paardekooper-mellema-2006}. Therefore the disc picks up a global vertical oscillation that lasts throughout the simulation. While this could potentially limit the comparison between the codes, we here assume that this oscillation does not couple to the buoyancy-induced modes. This means we can subtract the average vertical velocity from the \texttt{IDEFIX} results and compare the residuals with \texttt{FARGO3D}. We note that the importance of this oscillation will be weaker for larger planet masses, as it is tied to the initial conditions and how well the initial profile is resolved.

In Fig.~\ref{fig:local_vz_2d}, local analogs of the buoyancy wave fronts can be clearly seen \citep[see also][]{zhu-etal-2012}. It is also clear that \texttt{FARGO3D} shows a stronger response, resolving more modes, just as in the global case. This comes at the expense of structure on the grid scale, as can be seen from the stair-like patterns in the top panel. \texttt{IDEFIX} shows a smoother structure, possibly due its TVD nature. In the bottom panel, we show the result for an initial condition that has constant entropy $P/\rho^\gamma$, which eliminates any buoyancy response. While this setup does lead to non-zero vertical velocities, the amplitude is very much reduced and the characteristic pattern of the buoyancy response has almost vanished. Any residuals may be partly due to the fact that the numerical equilibrium is not exactly isentropic, and they can be further reduced by increasing the resolution. 

\begin{figure}
	\includegraphics[width=\columnwidth]{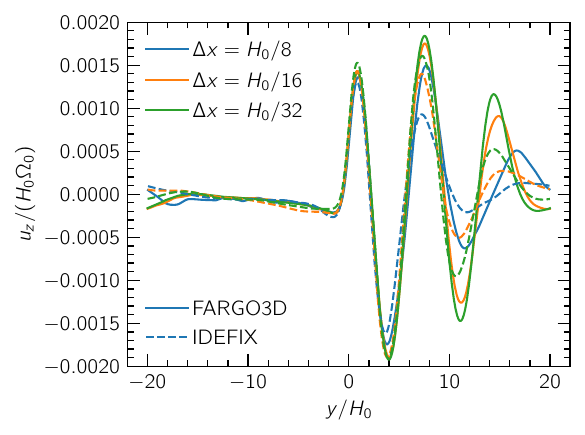}
	\caption{Vertical velocity at $\Omega_0 t=20$ at $z=H_0$ and $x=-H_0/2$ for the two codes at different resolutions. Initial conditions are isothermal, and $\nu=\chi=0$.}
	\label{fig:local_vz_resolution}
\end{figure}

The stronger buoyancy response in \texttt{FARGO3D} is further illustrated in Fig.~\ref{fig:local_vz_resolution}, where we show the vertical velocity as a function of $y$ at $z=H_0$ and $x=-H_0/2$. At the first maximum, the codes agree to within $10\%$ for all resolutions. The spread increases to $16\%$ at the first minimum, located around $y=4H_0$, and at the second maximum the spread is $50\%$, with the lowest resolution showing more damping, as expected. As time evolves, the buoyancy-induced pattern keeps extending, with more oscillations being added, but once an oscillation has appeared, its amplitude remains steady. We can use this to estimate the impact of the subtraction of the global vertical oscillation in \texttt{IDEFIX}. By looking at slightly different times, we probe different phases of the vertical oscillation, and because the buoyancy-induced pattern is steady, any difference in the measured amplitudes is due to the subtraction process. From this, we estimate that we can measure the amplitudes confidently to approximately $10\%$ accuracy. With this in mind, we can say that at the first maximum the codes agree at all resolutions, while significant differences start to arise at the first minimum. Comparing \texttt{FARGO3D} and \texttt{IDEFIX} for equal resolutions, we find a difference at the second maximum of $40\%$ for $\Delta x = H_0/8$, $20\%$ for $\Delta x = H_0/16$, and $13\%$ for $\Delta x = H_0/32$. It is plausible that if we would increase the resolution even further, \texttt{FARGO3D} and \texttt{IDEFIX} would agree at this second maximum to within our limit of $10\%$.    

\begin{figure}
	\includegraphics[width=\columnwidth]{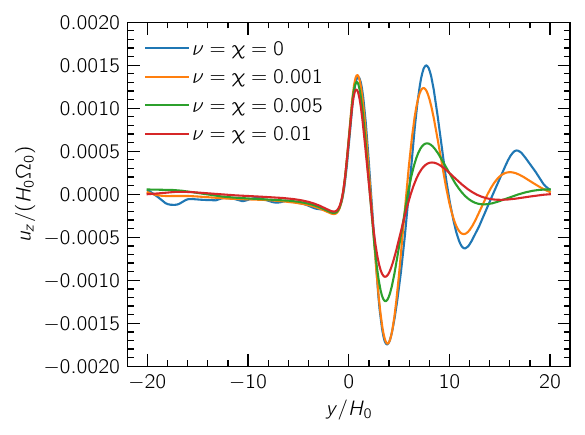}
	\caption{Vertical velocity at $\Omega_0 t=20$ at $z=H_0$ and $x=-H_0/2$ for different levels of diffusion (units of $\nu$ and $\chi$ are $H_0^2\Omega_0$). All results were obtained with \texttt{FARGO3D} at a resolution of $\Delta x = H_0/16$.}
	\label{fig:local_vz_diffusion}
\end{figure}

\subsection{Adding explicit diffusion}

We now consider cases where the diffusion coefficients are non-zero. For simplicity, we keep $\nu=\chi$, and therefore fix the Prandtl number to unity. This is purely for convenience, as in a protoplanetary disc it is usually the case that thermal diffusion (usually through radiation) dominates over viscous diffusion (through turbulence in the gas). Moreover, for the buoyancy response thermal diffusion is more important than viscous diffusion, and we found the results to be dominated by the value of $\chi$ rather than $\nu$. Nevertheless, we want to work towards a case where numerical diffusion is subdominant in the buoyancy response, and hence keep the viscosity coefficient in line with thermal diffusion. 

In Fig.~\ref{fig:local_vz_diffusion} we show the vertical velocity, again at $\Omega_0 t=20$, for different levels of diffusion. All results were obtained with \texttt{FARGO3D} at a resolution of $\Delta x = H_0/16$. Note that since the units of $\nu$ and $\chi$ are $H_0^2\Omega \approx \cs H_0$, the magnitude of $\nu$ in these units are the same as $\alpha$ in the usual $\alpha$ prescription for viscosity \citep{shakura-sunyaev-1973}. As expected, diffusion damps the buoyancy response, and while the first maximum is relatively unaffected, subsequent extrema are damped, to the extent that for $\nu=\chi=0.01~H_0^2\Omega_0$ there is only one more maximum established before the buoyancy response vanishes. Note also that the wiggly structure present in the inviscid run at $x<0$ is effectively removed even for modest levels of diffusion. 

Comparing Figs.~\ref{fig:local_vz_resolution}~and~\ref{fig:local_vz_diffusion}, we see that for our standard resolution of $\Delta x=H_0/16$, inviscid \texttt{IDEFIX} results are comparable to diffusive \texttt{FARGO3D} results with $\nu=\chi=0.001~H_0^2\Omega_0$. That is, for this specific problem setup (both numerical and physical), at this specific location in the domain, numerical diffusion in \texttt{IDEFIX} is similar to having $\nu=\chi=0.001~H_0^2\Omega_0$. At a lower resolution of $\Delta x = H_0/8$, the \texttt{IDEFIX} result is comparable to having $\nu=\chi=0.002~H_0^2\Omega_0$.

Perhaps a partial explanation for this behaviour lies in the ``total variation diminishing'' (TVD) nature of shock-capturing methods such as \texttt{IDEFIX} and \texttt{PLUTO}. The concept of TVD is related to shock handling and the avoiding of the Gibbs phenomenon. Methods that are TVD will not show artificial post-shock oscillations. However, flow structures subject to the background shear will lead to to an increase in TV, which is basically a measure of the integral of $|\partial W/\partial x|$, where $W$ is any flow variable. Hence, \texttt{IDEFIX} will damp strongly sheared structures such as the higher order buoyancy response, even though no shocks are present (the buoyancy response basically consists of density variations at constant pressure). \texttt{FARGO3D}, on the other hand, explicitly adds artificial viscosity whenever a convergent flow is detected \citep[see e.g.,][]{stone-norman-1992}. When such a flow is absent, as is largely the case of the buoyancy response, \texttt{FARGO3D} will not apply any artificial viscosity and the density structures are sheared out until the grid scale becomes very apparent, as can be seen in the top panel of Fig.~\ref{fig:local_vz_2d}.  

\begin{figure}
	\includegraphics[width=\columnwidth]{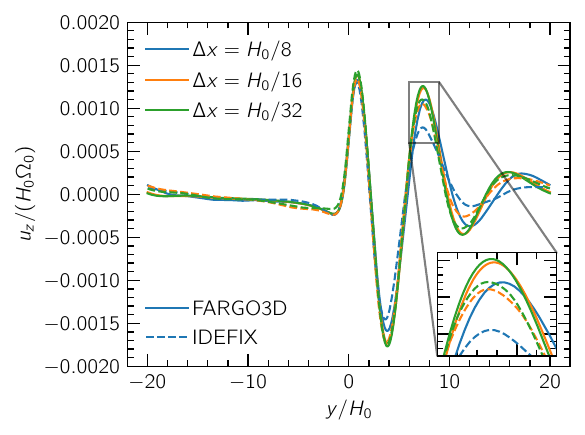}
	\caption{Vertical velocity at $\Omega_0 t=20$ at $z=H_0$ and $x=-H_0/2$ for $\nu=\chi=0.001~H_0^2\Omega_0$. Shown are results using \texttt{FARGO3D} (solid curves) and \texttt{IDEFIX} (dashed curves) at different resolutions.}
	\label{fig:local_vz_convergence}
\end{figure}

We now fix $\nu=\chi=0.001~H_0^2\Omega_0$, and compare \texttt{FARGO3D} and \texttt{IDEFIX} for different resolutions in Fig.~\ref{fig:local_vz_convergence}, where we again show the vertical gas velocity at $\Omega_0 t=20$, $x=-H_0/2$, and $z=H_0$. We first look at the codes individually, and look for signs of convergence with resolution. Since oscillations will be sheared apart more at larger $y$, they become more difficult to resolve and therefore a higher resolution is required. Focusing on the second maximum, the inset in Fig.~\ref{fig:local_vz_convergence} shows that both the solid curves (\texttt{FARGO3D}) and the dashed curves (\texttt{IDEFIX}) appear to converge to the same solution at our highest two resolutions. When comparing \texttt{FARGO3D} to \texttt{IDEFIX}, we need to keep in mind that the correction for the global vertical oscillation introduces an uncertainty of $\sim 10\%$. Within that uncertainty, both codes actually agree on the amplitude of the second maximum. This is consistent with our observation above, that the inviscid \texttt{IDEFIX} results looked similar to \texttt{FARGO3D} results with diffusion coefficients $\nu=\chi=0.001~H_0^2\Omega_0$. This agreement is of course what \emph{should} happen, in the regime where physical diffusion dominates over numerical diffusion, but this serves as a check that both codes are actually solving the same physical problem.

\begin{figure}
	\includegraphics[width=\columnwidth]{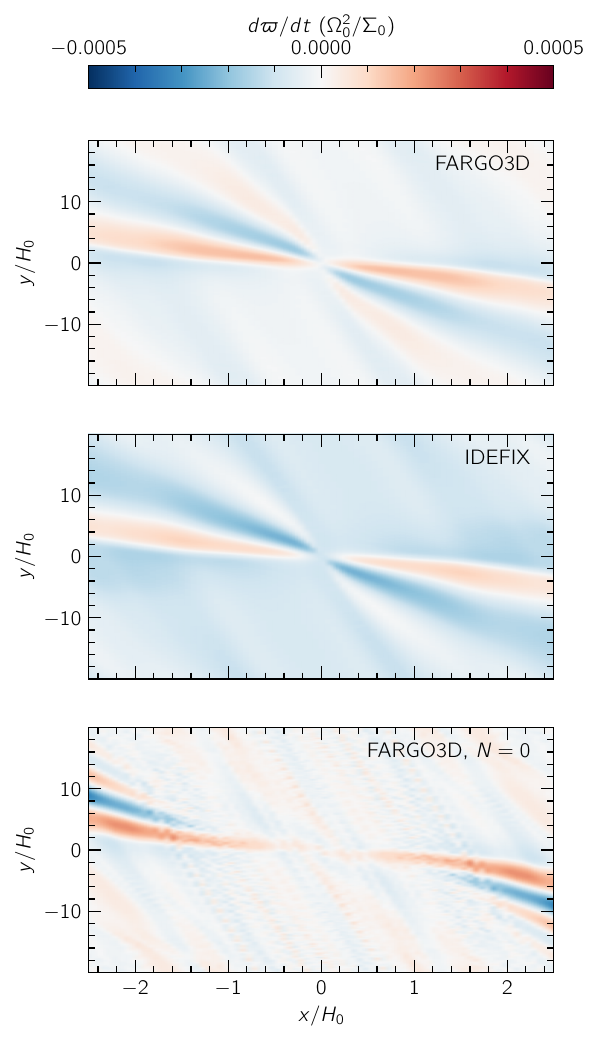}
	\caption{The vortensity source term, see Eq.~\eqref{eq:vortensity-source}. All simulations have $\nu=\chi=0$ and $\Delta x = H_0/16$, and results are shown at $\Omega_0 t=20$. Top panel: \texttt{FARGO3D}, middle panel: \texttt{IDEFIX}, bottom panel: \texttt{FARGO3D} with constant entropy initial condition.}
	\label{fig:local_vort_source}
\end{figure}

\subsection{Vortensity forcing}

We finally consider vortensity forcing in the local model, see Eq.~\eqref{eq:vortensity-source}. In Fig.~\ref{fig:local_vort_source}, we show heat maps of this source term for three different runs, all with $\nu=\chi=0$ at $\Omega_0 t=20$ with $\Delta x = H_0/16$. The vortensity source consists of the vortex stretching term $\left(\bm{\omega}_\rho\cdot\nabla\right)\vel$ and the baroclinic forcing $(\nabla\rho\times\nabla p)/\rho^3$. In the isentropic case (bottom panel of Fig.~\ref{fig:local_vort_source}), we found that the source is completely dominated by the vortex stretching term. This is expected, as this setup has constant entropy everywhere initially, and in the absence of thermal diffusion and shocks this means that entropy should remain constant at all times, rendering the $\nabla \rho \times \nabla P$ term identically zero. In reality, there is some numerical diffusion \citep[in particular since we do not use the entropy-based variant of \texttt{FARGO3D}, see][]{mcnally-etal-2019b}, as well as some converging flow structures that the code may interpret as shocks, which can lead to a non-zero vortensity source term. However, compared to the isothermal initial conditions, the resulting source is negligible. The stretching term mostly affects the spiral wave, with the corotation region largely unaffected.

In the top two panels of Fig.~\ref{fig:local_vort_source}, we see evidence for vortensity generation by the buoyancy response. We found that this is again almost completely due to the stretching term, as the baroclinic forcing term is very localized to the region close to the planet. This is in contrast to the global simulations, where also the baroclinic forcing correlates with the buoyancy response, albeit at a small level compared to the stretching term. The lack of baroclinic forcing of vortensity away from the planet in the local model is because there is no pressure gradient to create a large enough $\nabla \rho \times \nabla P$ term. The only horizontal pressure variations occur close to the planet (i.e., its atmosphere) and in the spiral wakes, which are essentially sound waves modified by rotation. The buoyancy response occurs at constant pressure, which strongly reduces the possibility of generating vortensity in the horseshoe region through baroclinic forcing. In the global models, on the other hand, there is a global radial pressure gradient, so that the density variations due to the buoyancy response combine with this radial pressure gradient to yield a non-zero $\nabla \rho \times \nabla P$ term. Both local and global models do agree that \texttt{FARGO3D} generates a stronger vortensity source than Riemann solvers such as \texttt{IDEFIX} and \texttt{PLUTO}, both through the vortex stretching term as well as the baroclinic forcing term.

\subsection{Section summary}

The local models have shown that the buoyancy response is damped more in \texttt{IDEFIX} compared to \texttt{FARGO3D}. In cases with explicit diffusion, both codes approach the same solution, but a detailed comparison is difficult because the \texttt{IDEFIX} solution includes a global vertical oscillation. Baroclinic forcing remains localized to a region very close to the planet because of the absence of a global pressure gradient, while vortex stretching is the main source of vortensity due to the buoyancy response.

\section{Discussion}
\label{sec:discussion}

In this section we discuss our results and their relevance to planet migration. We also note on the suitability of different codes to different physical problems. 

\subsection{Using $u_z$ as a proxy for the buoyancy torque}

A simplified analytical model by \citet{lubow-zhu-2014} showed that the width of the density ridges along buoyancy modes---which generate the related torque---depends on damping effects rather than pressure. This implies that a strict convergence test would require some explicit dissipation, as the density ridges would simply become narrower with increasing resolution otherwise, and then a comparison of torques at different resolutions and levels of dissipation. Given that the torque is only established after 3--4 libration timescales, or $\sim 200$--300 orbits, such a detailed comparison would be prohibitively expensive from a computational point of view.

To circumvent this, we used $u_z$ to measure the number and amplitude of buoyancy modes as a proxy for each code's ability to adequately capture the buoyancy response, and assume that the resulting torque will be proportional to the latter. The advantage of this method is that it only requires the buoyancy pattern to be established, which only takes $\sim 10$ planetary orbits.

We nevertheless expect that in our inviscid models, where the only source of diffusion is numerical dissipation, modes will be narrower in azimuth at higher resolution. We indeed observe this mainly for \texttt{PLUTO} and \texttt{IDEFIX} and between resolutions of 4--16 cells per scale height (cps), with the width of each undulation narrowing further very slightly at 32 cps (see Figs.~\ref{fig:uz-resolutions}~\&~\ref{fig:local_vz_resolution}). It is possible that the modes will not become infinitesimally thin due to nonlinear effects not predicted by the analytical model of \citet{lubow-zhu-2014}.

\subsection{How is planet migration affected?}
\label{sub:planet-migration}

\citet{zhu-etal-2012} first demonstrated that an embedded planet can excite vigorous vertical motion in the form of buoyancy modes, driving a torque with magnitude comparable to the total torque acting on the planet. \mcnally{} then showed how such a mechanism can drive a vortensity excess in the planet's corotating region, weakening the (outward) dynamical corotation torque \citep{mcnally-etal-2017} and speeding up inward migration. Their findings were limited to radiatively inefficient, quasi-adiabatic disks where cooling cannot damp these buoyancy oscillations, essentially maximizing their activity and showing that the planet can migrate inwards roughly twice as fast when this effect is considered. 

\citet{yun-etal-2022}, however, showed that including a treatment of radiation transport by means of cooling via a flux-limited diffusion approach \citep[see e.g.,][]{levermore-pomraning-1981} can significantly damp the buoyancy response when the timescale associated with the buoyancy frequency in Eq.~\eqref{eq:buoyancy-frequency} is comparable to the thermal diffusion timescale, suppressing it further as radiative diffusion becomes more efficient. This essentially quenches the buoyancy response for $z\gtrsim 2H$ even for optically thick disks at 1--2\,au for their opacity model \citep{bell-lin-1994}.

Their study suggests that radiative diffusion could greatly limit the radial and vertical range where buoyancy modes can operate and drive vortensity growth near a planet. As a result, the related torque might not actually act on the planet except for in the innermost au (Ziampras et al., in prep.) and \texttt{PLUTO}/\texttt{IDEFIX} can still be safely used for planet--disk interaction in 3D geometries without much meaningful loss in quality. 

\subsection{Which code to trust?}
\label{sub:which-code}

As we discussed in Sect~\ref{sub:different-numerics}, \texttt{PLUTO}/\texttt{IDEFIX} and \texttt{FARGO3D} use fundamentally different numerical approaches to solving Eqs.~\eqref{eq:navier-stokes}. \texttt{FARGO3D} uses second-order accurate spatial reconstruction with a 3-point stencil, but the timestepping scheme is formally a single-step, first-order, operator-split approach. Nevertheless, the staggered mesh allows conservation of momenta to machine precision, which is crucial for smooth astrophysical flows.

On the other hand, \texttt{PLUTO}/\texttt{IDEFIX} are at least second-order accurate in both time and space, and can capture shocks while properly conserving the relevant quantities (mass, momentum, total energy). However, they are susceptible to the high-Mach problem \citep{trac-pen-2004} which is present in highly supersonic flows such as protoplanetary disks. Essentially, more than 99\% of the total energy is kinetic, and therefore a strict conservation of total energy will cause any small errors during an energy update to propagate to the thermal energy. This means that even though total energy is conserved by construction, the ratio of kinetic to thermal energy will continuously change unpredictably, affecting the quality of results. This can be alleviated to an extent by instead conserving entropy \citep{ryu-etal-1993}, which we use for \texttt{PLUTO} models, however.

\begin{figure}
	\includegraphics[width=\columnwidth]{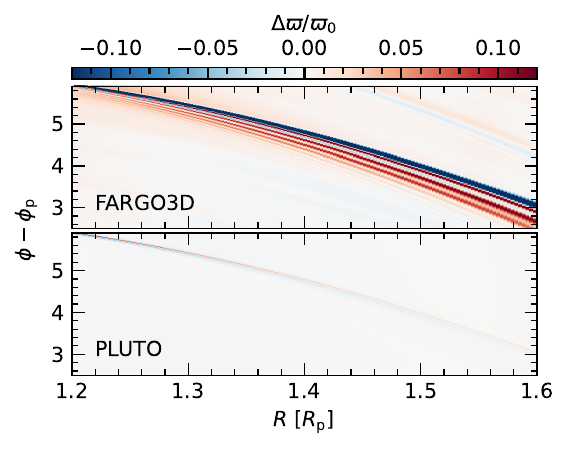}
	\caption{Perturbed vortensity around the spiral wake of the planet for vertically integrated, 2D models. The \texttt{PLUTO} solution shows perturbations only along the spiral shock as expected, while \texttt{FARGO3D} shows unphysical oscillations behind the shock front.}
	\label{fig:ringing}
\end{figure}

All in all, both codes have their shortcomings: \texttt{PLUTO}/\texttt{IDEFIX} are designed to capture shocks, while \texttt{FARGO3D} is best suited for linear, smooth, highly supersonic flows. Since this specific problem revolves around the excitation of linear waves, it makes sense that \texttt{FARGO3D} is more suitable here. When the focus is instead the planet-induced spiral shocks \citep{ogilvie-lubow-2002,rafikov-2002} it has been shown that even though both codes agree very well, the structure of a spiral shock will look smoother and more realistic in \texttt{PLUTO}/\texttt{IDEFIX}, while in \texttt{FARGO3D} one can see unphysical artifacts such as ``ringing'' behind the shock front (see Fig.~\ref{fig:ringing}). 

Of course, there exist problems where both codes are outshined by problem-specific approaches. When modeling hydrodynamic instabilities, for example, spectral codes such as \texttt{SNOOPY}\footnote{\url{https://ipag.osug.fr/\~lesurg/snoopy.html}} have been shown to perform very well at capturing delicate effects that occur on very small length scales comparable to the grid scale \citep[e.g.,][]{cui-latter-2022}.

\section{Summary}
\label{sec:summary}

We revisited the buoyancy response and the associated torque exerted on an embedded planet in an adiabatic protoplanetary disk \mcnallyp{}. We compared the Godunov-scheme codes \texttt{PLUTO} \citep{mignone-etal-2007} and \texttt{IDEFIX} \citep{lesur-etal-2023} to the finite-difference code \texttt{FARGO3D} \citep{benitez-llambay-etal-2016}.

We found that the two codes yield different results, and that the differences are inherent to the numerical schemes and not trivial. \texttt{FARGO3D} resolves more buoyancy modes, resulting in more efficient vortensity generation and a stronger negative torque, leading to a narrower, high-vortensity corotating region and faster inward migration. With \texttt{PLUTO}, the vortensity profile in the corotating region is significantly different and the buoyancy-induced torque is substantially weaker for our fiducial model.

We also found that the buoyancy response in \texttt{PLUTO} is strongly dependent on resolution and the numerical configuration used (solver, reconstruction order, timestepping order). With a sufficiently high-order (but also costly) numerical setup, \texttt{PLUTO} results resemble those of \texttt{FARGO3D} much more, implying that \texttt{FARGO3D} is more suitable for the study of this phenomenon.

Our local models confirmed these results, where we compared \texttt{FARGO3D} to \texttt{IDEFIX}. In this simplified geometry, again \texttt{FARGO3D} showed the stronger buoyancy response, keeping the density perturbations undamped even though they are taken apart by the background shear. \texttt{IDEFIX}, on the other hand, applies strong damping on structures close to the grid scale. In cases where we added explicit diffusion, both codes agreed on the solution at high enough resolution, with measured differences $<10\%$. In the local models, the baroclinic forcing term vanishes away from the planet due to the lack of a global pressure gradient, which will lead to different evolution of the vortensity in the horseshoe region compared to global models.

We highlight, however, that the buoyancy-related torque might not be as present in realistic, radiative disks \citep{yun-etal-2022}. This is heavily dependent on the disk model, and we will revisit this in future work. Nevertheless, given the niche of this effect, and the fact that both codes have been used successfully for planet--disk interaction and agree with each other in a wide variety of problems, we expect that this will not affect the codebase of finite-volume codes. Our intent is simply to raise awareness to the fact that sometimes one numerical approach will be more suitable than another for a particular problem. 

\section*{Acknowledgements}
AZ would like to thank Mario Flock for very helpful comments and fruitful discussions. This research utilized Queen Mary's Apocrita HPC facility, supported by QMUL Research-IT (http://doi.org/10.5281/zenodo.438045). This work was performed using the DiRAC Data Intensive service at Leicester, operated by the University of Leicester IT Services, which forms part of the STFC DiRAC HPC Facility (www.dirac.ac.uk). The equipment was funded by BEIS capital funding via STFC capital grants ST/K000373/1 and ST/R002363/1 and STFC DiRAC Operations grant ST/R001014/1. DiRAC is part of the National e-Infrastructure. AZ and RPN are supported by STFC grant ST/P000592/1, and RPN is supported by the Leverhulme Trust through grant RPG-2018-418. This project has received funding from the European Research Council (ERC) under the European Union’s Horizon 2020 research and innovation programme (grant agreement No 101054502). All plots in this paper were made with the Python library \texttt{matplotlib} \citep{hunter-2007}.

\section*{Data Availability}

Data from our numerical models are available upon reasonable request to the corresponding author.

\bibliographystyle{mnras}
\bibliography{refs}

\bsp	
\label{lastpage}
\end{document}